\documentstyle[11pt,epsf]{article}
\begin{document}
\setcounter{page}{1}
\pagestyle{plain}
\setcounter{equation}{0}
%
%
%
\ \\[12mm]
\begin{center}
    {\bf DETERMINISTIC EXCLUSION PROCESS WITH A STOCHASTIC DEFECT:\\[1mm]
         MATRIX-PRODUCT GROUND STATES}
\\[20mm]
\end{center}
\begin{center}
\normalsize
	Haye Hinrichsen$^{\diamond,}
	\footnote{e-mail: \tt fehaye@wicc.weizmann.ac.il}$ and
        Sven Sandow$^{\star,}
	\footnote{e-mail: \tt sandow@dds.phys.vt.edu}$ \\[13mm]
	$^{\diamond}$ {\it Department of Physics of Complex Systems\\
	Weizmann Institute of Science\\ Rehovot 76100, Israel}\\[4mm]
	$^{\star}$ {\it Department of Physics and Center for Stochastic
	Processes in Science and Engineering\\
	Virginia Polytechnic Institute and State University\\
	Blacksburg, VA 24061-0435, USA}
	\end{center}
\vspace{3cm}
{\bf Abstract:} We study a one-dimensional anisotropic
exclusion model describing particles moving
deterministically on a ring with a single defect across
which they move with probability $0 < q < 1$.
We show that the stationary state of this model can be 
represented as a matrix-product state.
\vspace{31mm}
\\[3mm]
\rule{7cm}{0.2mm}
\begin{flushleft}
\parbox[t]{3.5cm}{\bf Key words:}
\parbox[t]{12.5cm}{Reaction-diffusion models, 
		   asymmetric exclusion process,\\
	           matrix-product states}
\\[1mm]
\parbox[t]{3.5cm}{\bf PACS numbers:} {02.50.Ey, 05.60.+w, 05.70.Ln}
%
%
\parbox[t]{12.5cm}{}
\end{flushleft}
\normalsize
\thispagestyle{empty}
\mbox{}
\pagestyle{plain}
%
%
%
\newpage
\section{Introduction}
\setcounter{page}{1}
\setcounter{equation}{0}
%
%
%
The one-dimensional asymmetric exclusion process (ASEP) is a lattice model
which describes particles hopping in a preferred direction with
stochastic dynamics and hard core exclusion. It has been used 
to describe various phenomena like growth processes \cite{Diffusive,Growth},
polymerization \cite{Polymers}, and traffic flow \cite{Traffic}.
Although the ASEP is one of the simplest stochastic many-particle 
models it shows a rich behavior ranging from phase transitions
\cite{ddm,PhaseTransitions} to the formation of shocks \cite{Shocks}.
During the last few years exact solutions have been found for the stationary
under various boundary conditions. 
In particular the ASEP with random sequential dynamics
and external particle input and output at the 
boundaries has been analyzed in 
detail~\cite{PhaseTransitions,Derrida,sa,Algebra}. There are only
few exact results for dynamic properties (relaxational behavior) 
of the ASEP \cite{DynamicASEP}.

%
%
A mathematical tool which has proved to be very useful for the
study  of stochastic lattice models is
the so-called matrix-product  state (MPS) technique 
\cite{HakimEtAl}-\cite{Honecker},\cite{Derrida}-\cite{Algebra}. 
Matrix-product states can be understood as a generalization
of ordinary factorizable states with a product
measure in which numbers are replaced by non-commutative object.
This allows the original problem to be reformulated in terms of algebraic
relations. The main advantage of this technique is that once a nontrivial 
representation of the algebra is available, physical quantities like 
density-density correlation functions can be computed very easily. 
In 1993 Derrida\ {\it et.\ al.\ }showed that the stationary state of the
ASEP with external particle input and output can be written as a MPS.
Since then matrix-product methods have been applied to various other
problems, e.g. to three-state models~\cite{ThreeState,MovingDefects}, 
excited states in the ASEP \cite{Dynamics}, certain reaction-diffusion 
models~\cite{ReactionDiffusion} and models with quasi-parallel updates
\cite{MatrixParallel}-\cite{Honecker}. 
Although a lot of work on MPS's has been done,
the full range of applications is not yet known\footnote{
As it has been shown recently,  for every 
one-dimensional stochastic lattice model with local random 
sequential dynamics  the eigenstates of the 
Hamiltonian can be written as MPS's \cite{Sandow}. 
However, matrix representations which are useful for practical
purposes seem to be limited to a few models.}.

%
%
So far the matrix-product technique has been applied mostly to systems
which are homogeneous in the bulk. It is therefore interesting to
investigate whether this method can also be applied to systems with impurities.
In case of the ASEP two kinds of impurities are discussed in the literature.
The first kind of impurities are {\em defective particles} which jump
with a rate less than that of other particles 
\cite{MovingDefects,Mallick,BoseEinstein}. 
Such 'moving' impurities can be visualized 
as slow cars on a motor way which (for sufficiently high particle
density) induce traffic jams - a phenomenon which has been related recently to
Bose-Einstein condensation \cite{BoseEinstein}.
To this problem the matrix-product technique
has been applied successfully~\cite{MovingDefects,Mallick}.
The second kind of impurities are {\em stationary defects} which can
be introduced by lowering the hopping rate at specific bonds.
As in the previous case, such impurities can provoke the formation
of shocks \cite{StationaryShocks}.
Defects of this kind play an important role in traffic flow
applications wherefore exact solutions are particularly interesting.
However, for some reason ASEP's with
stationary defects are much harder to solve
than systems with moving impurities. Exact results exist only in case of an
exclusion model with parallel dynamics and deterministic hopping
in the bulk \cite{ParallelUpdate}-\cite{Blockage}. The case of
stochastic hopping or random sequential dynamics has not yet been
solved exactly. Moreover matrix-product techniques have not been
applied to this type of problems.

%
%
In this paper we consider the stationary state of the ASEP
on a ring with various kinds of parallel dynamics, deterministic hopping
in the bulk and a single stationary defect. Using Bethe-Ansatz
techniques, this problem was first solved in Ref. \cite{Blockage}
in case of alternating parallel updates on two sublattices.
The solution, however, is fairly complicated
and amounts to a list of rules for an operative construction
of the stationary state. In the present work we solve the problem
in a more compact  way by using the matrix-product technique.
Moreover this formalism allows models with different update sequences 
to be solved within the same framework. In order to demonstrate this, 
we consider two different update sequences .
Our main intention, however, is to outline a method which might help to
solve the problem of stochastic hopping in presence of a defect.

%
%
The paper is organized as follows. In Sect. \ref{ModelDefinition} we
define the ASEP on a ring with a defect and introduce two different dynamical
rules. In Sect. \ref{FourStateModel} we show that the two-state model on a
ring can be reformulated as a four-state model on a linear chain. 
Using this mapping we formulate 
a matrix-product ansatz (see Sect.
\ref{AlgebraSection}) leading to a set of algebraic equations
which turns out to be the same for both update sequences.
In Sect. \ref{RepresentationSection} we give simple two-dimensional
representations of this algebra. Using these representations, some
physical quantities are derived and discussed in Sect. \ref{PhysicalSection}.
Finally in Sect. \ref{ConclusionSection} we summarize our results and
discuss possible generalizations.


\section{The model}
\label{ModelDefinition}

%
%
The exclusion model we are going to study is defined as follows:
Particles of one species move in clockwise direction on a one-dimensional ring
with an even number of sites $L=2N$. Each site can be either  free or
occupied by one particle.
The time evolution is discrete and defined by a certain sequence
of parallel updates to be specified below. In the bulk of the chain 
these updates are deterministic, i.e. in each time step the particles
move forward provided that the following site is empty. Between
two particular sites (by convention sites $L$ and $1$) a defective bond
is inserted where the particles hop stochastically with a given
probability $0 < q < 1$ (see Fig. 1).

\begin{figure}
\begin{center}~
\epsfxsize=110mm
\epsffile[18 550 593 774]{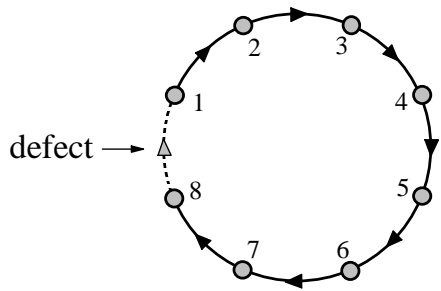}
\end{center}
\label{FigureRing}
\caption{\it The ASEP on a ring with $L=2N=8$ sites. In the bulk particles
		 hop deterministically in clockwise direction. A defect is
         	 introduced between sites $L$ and $1$ where particles hop with
		 probability~$q$.}
\end{figure}

%
%
In what follows we will consider two different update sequences. The first
one we call {\em symmetrized sequential dynamics}:
In each time step the first update takes place at the sites 
$(N,N+1)$ just opposite the defect. Then a sequence of pairwise
updates follows. These updates are arranged symmetrically with respect
to the defect. The first pair to be updated is $(N-1,N)$ and $(N+1,N+2)$,
followed by $(N-2,N-1)$ and $(N+2,N+3)$ up to $(1,2)$ and $(L-1,L)$. The last
pair of sites to be updated is the defect $(L,1)$. As we will see below,
this is the simplest dynamical rule in terms of the matrix formalism.
\\ \indent
The other dynamical rule we will use is called
{\em sublattice-parallel dynamics}. Here we have to assume that
$N=L/2$ is an even number. Each time step consists of two separate
half time steps.  In the first half time step
particles located at odd sites move one step in clockwise direction provided
that the next site is empty.  Then in the second half time step particles
at even sites move forward in the same way except for the defect where
such moves take place with probability $q$. This dynamical rule was 
introduced in Ref. \cite{Blockage} for solving the deterministic ASEP 
with a defect. A similar dynamics was also used in
Refs. \cite{ParallelUpdate,MatrixParallel} in case of
the ASEP with external particle input and output.

%
%
Symmetrized sequential and sublattice-parallel dynamics
can be cast in a more formal way as follows.
Let $\tau_j=0,1$ be the occupation number at site $j$ and consider the
space of all configurations in a canonical configuration basis. 
Let $\cal T$ (${\cal T}^{(q)}$) be the two-site hopping matrix in the 
bulk (at the defect):
\begin{equation}
{\cal T} \;=\; \left(
\begin{array}{cccc}
1 & 0 & 0 & 0 \\
0 & 1 & 1 & 0 \\
0 & 0 & 0 & 0 \\
0 & 0 & 0 & 1
\end{array} \right)\,,
\hspace{10mm}
{\cal T}^{(q)} \;=\; \left(
\begin{array}{cccc}
1 & 0 & 0 & 0 \\
0 & 1 & q & 0 \\
0 & 0 & 1-q & 0 \\
0 & 0 & 0 & 1 
\end{array} \right)\,.
\end{equation}
Then the transfer matrix $T_{seq}$ for symmetrized sequential dynamics reads
\begin{equation}
\label{SequentialTransferMatrix}
T_{seq} \;=\; {\cal T}^{(q)}_{L,1}\,({\cal T}_{1,2}{\cal T}_{L-1,L})\,
({\cal T}_{2,3}{\cal T}_{L-2,L-1}) \, \ldots \, 
({\cal T}_{N-1,N}{\cal T}_{N+1,N+2})\,{\cal T}_{N,N+1}\,,
\end{equation}
where the indices indicate pairs of sites on which ${\cal T}$ acts.
The transfer matrix for sublattice-parallel updates 
factorizes into two factors 
\begin{equation}
T_{par}\;=\; T^{(2)}_{par}\,T^{(1)}_{par}
\end{equation}
which are given by
\begin{eqnarray}
\label{SublatticeTransferMatrix}
T^{(1)}_{par} &=&
			({\cal T}_{1,2}{\cal T}_{L-1,L})\,
            ({\cal T}_{3,4}{\cal T}_{L-3,L-2})\, \ldots
            ({\cal T}_{N-1,N}{\cal T}_{N+1,N+2})\,, 
\\
T^{(2)}_{par} &=& 
	    {\cal T}^{(q)}_{L,1}\, ({\cal T}_{2,3}{\cal T}_{L-2,L-1})\,
	    ({\cal T}_{4,5}{\cal T}_{L-4,L-3})\, \ldots
	    ({\cal T}_{N-2,N-1}{\cal T}_{N+2,N+3})\, {\cal T}_{N,N+1}  \,.
		   \nonumber
\end{eqnarray}
Notice that both transfer matrices $T_{par}$ and $T_{seq}$
differ only in the sequence of their (noncommutative) factors.
There are, however, many other transfer matrices with different
update sequences. As we will see below, the matrix-product formalism
allows to compute ground states for various dynamical rules
from the same matrix representation just by rearranging the
factors in the matrix product. This is an important advantage
of the matrix-product method compared to direct techniques as presented
in Ref. \cite{Blockage}.


\section{Reformulation as a four state model}
\label{FourStateModel}

\begin{figure}
\begin{center}~
\epsfxsize=120mm
\epsffile[18 620 577 780]{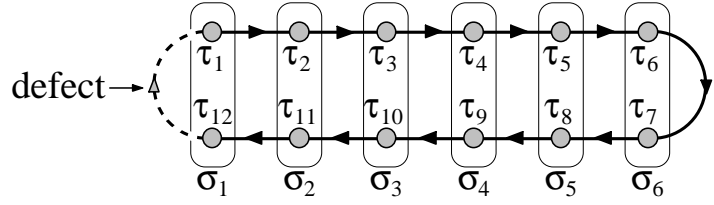}
\end{center}
\caption{\it By grouping pairs of sites, the two-state ASEP on a ring can be
	 regarded as a four-state model on a linear chain.}
\label{FigureGroups}
\end{figure}

%
%
Although the defect breaks translational invariance, the model is still
symmetric under reflections with respect to the origin combined with
particle-hole symmetry. Since we expect correlations in the stationary state
to be subject to this symmetry, it is natural to introduce a matrix ansatz 
 which exploits this symmetry. This can be done by 
grouping pairs of sites together which are located symmetrically 
with respect to the defect. This mapping defines a four-state model 
on a linear chain with $N$ sites and closed boundaries 
which is equivalent to the original one (see Fig. \ref{FigureGroups}).
In this language the two-site hopping matrix ${\cal T}^{(q)}_{L,1}$ in the
original notation is equivalent to an one-site operator ${\cal L}_1$ at the
left boundary acting in a four-dimensional space. Similarly ${\cal T}_{N,N+1}$
corresponds to an one-site operator ${\cal R}_N$ at the right boundary.
The block-spin variables $\sigma_i$ = $(\tau_{L-i+1}\,,\,\tau_i)$ can take
four different values $\{0,1,2,3\} = \{(0,0),(0,1),(1,0),(1,1)\}$. In
this basis the boundary hopping matrices read
\begin{equation}
\cal{L}=\left(
\begin{array}{cccc}
1 & 0 & 0 & 0 \\
0 & 1 & q & 0 \\
0 & 0 & 1-q & 0 \\
0 & 0 & 0 & 1 
\end{array} \right) \,,
\hspace{10mm}
\cal{R}=\left(
\begin{array}{cccc}
1 & 0 & 0 & 0 \\
0 & 0 & 0 & 0 \\
0 & 1 & 1 & 0 \\
0 & 0 & 0 & 1 
\end{array} \right) \,.
\end{equation}

\noindent
In the bulk pairs of opposite hopping matrices
$({\cal T}_{i,i+1}\,{\cal T}_{L-i,L-i+1})$ are grouped together resulting
in a two-site operator ${\cal S}_{i,i+1}$ which is given by
a $16 \times 16$ matrix:
\begin{equation}
\cal{S}={\small \left(
\begin{array}{cccc|cccc|cccc|cccc}
1&0&0&0 & 0&0&0&0 & 0&0&0&0 & 0&0&0&0 \\
0&1&0&0 & 1&0&0&0 & 0&0&0&0 & 0&0&0&0 \\
0&0&0&0 & 0&0&0&0 & 0&0&0&0 & 0&0&0&0 \\
0&0&0&0 & 0&0&0&0 & 0&0&0&0 & 0&0&0&0 \\
\hline
0&0&0&0 & 0&0&0&0 & 0&0&0&0 & 0&0&0&0 \\
0&0&0&0 & 0&1&0&0 & 0&0&0&0 & 0&0&0&0 \\
0&0&0&0 & 0&0&0&0 & 0&0&0&0 & 0&0&0&0 \\
0&0&0&0 & 0&0&0&0 & 0&0&0&0 & 0&0&0&0 \\
\hline
0&0&1&0 & 0&0&0&0 & 1&0&0&0 & 0&0&0&0 \\
0&0&0&1 & 0&0&1&0 & 0&1&0&0 & 1&0&0&0 \\
0&0&0&0 & 0&0&0&0 & 0&0&1&0 & 0&0&0&0 \\
0&0&0&0 & 0&0&0&0 & 0&0&0&1 & 0&0&1&0 \\
\hline
0&0&0&0 & 0&0&0&0 & 0&0&0&0 & 0&0&0&0 \\
0&0&0&0 & 0&0&0&1 & 0&0&0&0 & 0&1&0&0 \\
0&0&0&0 & 0&0&0&0 & 0&0&0&0 & 0&0&0&0 \\
0&0&0&0 & 0&0&0&0 & 0&0&0&0 & 0&0&0&1 
\end{array} \right)}     \,.
\end{equation}
In terms of these operators, the transfer matrix for sequential 
dynamics (\ref{SequentialTransferMatrix}) takes the simple form
\begin{equation}
\label{tsequ}
T_{seq} \;=\; {\cal L}_1\, {\cal S}_{1,2}\,{\cal S}_{2,3}\,\ldots\,
{\cal S}_{N-1,N}\, {\cal R}_N     \,.
\end{equation}
Similarly the transfer matrix for symmetrized sublattice-parallel 
dynamics (\ref{SublatticeTransferMatrix}) can be written as
\begin{equation}
T^{(1)}_{par} = {\cal S}_{1,2}\,{\cal S}_{3,4}\,\ldots\,{\cal S}_{N-1,N} \,,
\hspace{10mm}
T^{(2)}_{par} = {\cal L}_1\, {\cal S}_{2,3}\,{\cal S}_{4,5}\,\ldots\,
	       {\cal S}_{N-2,N-1}\, {\cal R}_N   \,.
\end{equation}
So far, we have only rewritten the original model on a ring as a four 
state model on a linear chain.
%
%
%
%
%
\section{The matrix-product formalism} 
\label{AlgebraSection}
As we will show, the four-state formulation allows to write
the stationary state of the asymmetric exclusion model with a
defect as a matrix-product state. Let us define noncommutative 
operators (matrices) $A_0,A_1,A_2,A_3$ and $B_0,B_1,B_2,B_3$ as well as
bra and ket vectors $\langle W|$ and $|V\rangle$ in some auxiliary space
and demand that they obey the relations
\begin{eqnarray}
\label{BulkShift}
{\cal S}\,(\vec{A} \otimes \vec{B}) &=& (\vec{B} \otimes \vec{A})\\
\label{RightBC}
{\cal R}\, \vec{A} \, |V\rangle &=& \vec{B} \, |V\rangle\,\\
\label{LeftBC}
\langle W| \, {\cal L}\, \vec{B}  &=& \langle W|\,\vec{A}
\;\;.\end{eqnarray}
where $\vec{A}=(A_0,A_1,A_2,A_3)$, $\vec{B}=(B_0,B_1,B_2,B_3)$ and
'$\otimes$' denotes the tensor product in configuration space.
This algebra can be used to construct the ground state of the stochastic model
we are studying. Let us first consider the case of symmetrized sequential
dynamics. As we will show below, the probability
$P_{seq}(\sigma_1,\ldots,\sigma_N)$ to find the stationary system
in the configuration $\{\sigma_1,\ldots,\sigma_N\}$ can be written as
\begin{equation}
\label{SequMatrixProductComplete}
P_{seq}(\sigma_1,\ldots,\sigma_N) \;=\; \frac{1}{Z^{seq}_N}\,
\langle W| A_{\sigma_1} A_{\sigma_2}\ldots  A_{\sigma_N} | V \rangle\,,
\end{equation}
where $Z^{seq}_N$ is a normalization
constant which is given by
\begin{equation}
Z^{seq}_N\;=\; \langle W|C^N|V \rangle\,
\hspace{10mm}
\mbox{with}
\hspace{10mm}
C\;=\; A_0+A_1+A_2+A_3\,.
\end{equation}
For example, the probability to find the four-state chain with $N=4$ sites
in the configuration $\{1,0,3,2\} = \{(0,1),(0,0),(1,1),(1,0)\}$ is given
by $P_{seq}(1,0,3,2)=\langle W | A_1 A_0 A_3 A_2 | V \rangle /
 \langle W | C^4  | V \rangle$. 
 
The mechanism ensuring the stationarity of these probabilities
is precisely the one introduced in Ref. \cite{Rajewsky} and works as follows:
Formally Eq.~(\ref{SequMatrixProductComplete})
may be written as $|P_{seq}\rangle=
({Z^{seq}_N})^{-1}\,\langle W| \vec{A}^{\otimes N} |V\rangle$.
Applying the transfer matrix
$T_{seq} \;=\; {\cal L}_1\, {\cal S}_{1,2}\,{\cal S}_{2,3}\,\ldots\,
{\cal S}_{N-1,N}\, {\cal R}_N$ to this state the matrix ${\cal R}$
at the right end of the chain generates the vector
$\vec{B}=(B_0,B_1,B_2,B_3)\;$ at the $N$-th position in the product 
$\vec{A}^{\otimes N}$ (cf. Eq. (\ref{RightBC})). Then, by successively
applying ${\cal S}$, the generated vector
$\vec{B}$ is commuted to the left (cf. Eq. (\ref{BulkShift})).
Finally, when reaching the left boundary, the vector $\vec{B}$ is 
turned into $\vec{A}$
by the action of the defect hopping matrix ${\cal L}$ 
(cf. Eq. (\ref{LeftBC})). Consequently, the application of
the transfer matrix $T_{seq}$ on the state $|P_{seq}\rangle$ 
results in the same state $|P_{seq}\rangle\;$.
Thus the state $|P_{seq}\rangle$ is a stationary state of the
transfer matrix (\ref{tsequ}), i.e. of the transfer matrix
(\ref{SequentialTransferMatrix}).

For sublattice-parallel dynamics we have to use a slightly different ansatz
which involves alternating matrices. We write the stationary state as
\begin{equation}
\label{ParMatrixProductComplete}
P_{par}(\sigma_1,\ldots,\sigma_N) \;=\; \frac{1}{Z^{par}_N}\,
\langle W| \, (A_{\sigma_1} B_{\sigma_2})\,( A_{\sigma_3} B_{\sigma_4})
		   \ldots  (A_{\sigma_{N-1}} B_{\sigma_N})\, | V \rangle\,,
\end{equation}
where
\begin{equation}
Z^{par}_N\;=\;\langle W| \,(\sum_{i=1}^4 A_i \,\sum_{j=1}^4 B_j)^{N/2}
\,| V \rangle\,.
\end{equation}
For this type of dynamics the mechanism ensuring the stationarity 
of the probabilities (\ref{ParMatrixProductComplete}) is precisely the one
proposed in Ref. \cite{MatrixParallel}.  Again we use use the
algebraic relations (\ref{BulkShift})-(\ref{LeftBC}).
Writing $|P_{par}\rangle = ({Z^{par}_N})^{-1}\,\langle W| \,
(\vec{A}\otimes \vec{B})^{\otimes N/2}\, |V\rangle$
one can easily verify that in each half time step the vectors
$\vec{A}$ and $\vec{B}$ are exchanged:
\begin{eqnarray}
T_{par}^{(2)} [(\vec{A}\otimes \vec{B})^{\otimes N/2}] &=&
	       (\vec{B}\otimes \vec{A})^{\otimes N/2}\,,\\
\langle W |
T_{par}^{(1)} [(\vec{B}\otimes \vec{A})^{\otimes N/2}] 
| V \rangle &=& \langle W |
		(\vec{A}\otimes \vec{B})^{\otimes N/2} | V \rangle\,.
\end{eqnarray}
Therefore the state $|P_{par}\rangle$ is invariant under application
of the transfer matrix $T_{par}$.

It should be emphasized that
both symmetrized sequential and sublattice-parallel dynamics use the same
algebra (\ref{BulkShift})-(\ref{LeftBC}). One can also consider various other
update sequences leading to (\ref{BulkShift})-(\ref{LeftBC}).
For example, if updating is started in the middle of the system
and proceeds to both ends from there, the stationary state
can be written as $\langle W| \,
\vec{A}^{\otimes N} \vec{B}^{\otimes N}\, |V\rangle\;$ with
the $A_i\,,\;B_i$ and $\langle W|\,,\;|V \rangle$ obeying
(\ref{BulkShift})-(\ref{LeftBC}).
In other words, having determined a nontrivial representation of the algebra
given above, one can immediately compute stationary states for various
update sequences.
%
%
%
\section{Representation of the algebra}
\label{RepresentationSection}
Eqs. (\ref{BulkShift})-(\ref{LeftBC}) define an algebra of eight objects
$A_0, A_1, A_2, A_3, B_0, B_1, B_2, B_3$. There are sixteen bulk equations
\begin{eqnarray}
\label{BulkEquations}
&&[A_i,B_i]\;=\;0 \hspace{50mm} (i=0\ldots 3) \nonumber \\[2mm] 
&&B_0A_2=B_0A_3=B_1A_0=B_1A_2=B_1A_3=B_3A_0=B_3A_2\;=\;0 \\[2mm]
&&A_iB_j+A_jB_i\;=\;B_iA_j \hspace{34mm} \nonumber
(i,j)\in\{(0,1),(2,0),(2,3),(3,1)\}\\[2mm]
&&A_0B_3+A_1B_2+A_2B_1+A_3B_0\;=\;B_2A_1\,, \nonumber
\end{eqnarray}
four equations at the defect (left boundary of the four-state model)
\begin{eqnarray}
\label{LeftBoundary}
\langle W| B_0 &=& \langle W| A_0 \nonumber \\
\langle W| (B_1+q B_2) &=& \langle W| A_1\\
\langle W| (1-q) B_2 &=& \langle W|A_2\nonumber \\
\langle W| B_3 &=& \langle W| A_3\,, \nonumber
\end{eqnarray}
and four equations opposite of the defect (right boundary):
\begin{eqnarray}
\label{RightBoundary}
A_0 |V\rangle &=& B_0|V\rangle \nonumber \\
0 &=& B_1 |V\rangle \\
(A_1+A_2) |V\rangle &=& B_2 |V\rangle \nonumber \\
A_3|V\rangle  &=& B_3 |V\rangle \,. \nonumber
\end{eqnarray}
Notice that the algebra is invariant under the replacement 
\cite{ThanksToAndreas}
\begin{equation}
\label{Freedom}
A_0 \rightarrow \lambda\,A_0\,, \hspace{10mm}
B_0 \rightarrow \lambda\,B_0\,, \hspace{10mm}
A_3 \rightarrow \lambda^{-1}\,A_3\,, \hspace{10mm}
B_3 \rightarrow \lambda^{-1}\,B_3\,, \hspace{10mm}
\end{equation}
where $\lambda$ is some number.

The algebra (\ref{BulkEquations})-(\ref{RightBoundary})  
has a complex structure and thus it is practically
impossible to determine its representations directly. In order to simplify
the problem, we therefore use the following ansatz:
\begin{equation}
\label{SimplificationAnsatz}
B_0=A_0 \,, \hspace{10mm}
B_1=A_1 - \mbox{\bf 1} \,, \hspace{10mm}
B_2=A_2 + \mbox{\bf 1} \,, \hspace{10mm}
B_3=A_3 \,. \hspace{10mm}
\end{equation}
A similar ansatz has been used in Refs. \cite{Rajewsky,Honecker}
in order to relate sublattice-parallel and random-sequential updates
in the ASEP with external particle input and output. It is obvious that
Eq.~(\ref{SimplificationAnsatz}) imposes strong constraints and
therewith reduces the space of solutions. However, it turns out that
it still includes nontrivial physical representations.

Inserting Eq.~(\ref{SimplificationAnsatz}) into 
Eqs.~(\ref{BulkEquations})-(\ref{RightBoundary}),
we obtain a reduced algebra of four objects. It consists of seven
bulk equations
\begin{equation}
\label{SimpleBulkAlgebra}
A_1A_0 = A_0\,, \hspace{10mm} A_1A_3 = A_3\,, \hspace{10mm} A_1A_2 = A_2
\,, \hspace{10mm} A_0A_2 = A_3A_2 = A_0A_3 = A_3A_0 = 0
\end{equation}
and two boundary conditions:
\begin{equation}
\label{SimpleBoundaryConditions}
\langle W|A_2=\frac{1-q}{q}\langle W|\,,
\hspace{20mm}
A_1|V\rangle = |V\rangle\,.
\end{equation}
This algebra is much simpler and can be analyzed systematically
on a computer. In fact, we found the following two-dimensional
representation:
$$
A_0=\lambda\left(\begin{array}{cc}0&1\\0&1\end{array}\right)\,, \hspace{7mm}
A_1=\left(\begin{array}{cc}1&0\\0&1 \end{array}\right)\,, \hspace{7mm}
A_2=\frac{1-q}{q}
	\left(\begin{array}{cc}1&1\\0&0\end{array}\right)\,, \hspace{7mm}
A_3=0\,,
$$ \begin{equation}
\label{Representation}
\langle W|=(1,1)\,, \hspace{15mm}
| V \rangle = \left( \hspace{-2mm}
\begin{array}{c} 1 \\[-1mm] 1 \end{array} \hspace{-1.5mm}\right)\,,
\end{equation}
where $\lambda$ is a free parameter. Algebraically $\lambda$ is related
to the invariance of the algebra in~Eq. (\ref{Freedom}). Physically
it is related to the conservation of the number of particles. In fact,
as will be discussed in the next section,
the ground states (\ref{SequMatrixProductComplete}) and
(\ref{ParMatrixProductComplete}) describe grand-canonical ensembles of systems
with different particle numbers where the parameter $\lambda$ plays the role
of a fugacity  \cite{ThanksToAndreas}.

As can be verified easily, the canonical ensemble described by the above 
representation includes {\it all} sectors with $M \leq N=L/2$ particles 
(if there were more than $L/2$ particles, at least one of the block spins
$\sigma_i$ = $(\tau_{L-i+1}\,,\,\tau_i)$ would be in the state (1,1). Because
of $A_3=0$ this implies that the corresponding matrix product vanishes). 
However, the algebra
(\ref{SimpleBulkAlgebra})-(\ref{SimpleBoundaryConditions}) is invariant
under the exchange of the matrices $A_0 \leftrightarrow A_3$ which
immediately gives a second set of representations for systems with
more than $L/2$ particles. Physically this invariance is related to
the particle-hole symmetry in the ASEP.
%
%
%
%
%
\section{Some physical quantities for the model with 
symmetrized sequential dynamics}
\label{PhysicalSection}
\subsection{Micro-canonical ensemble:
Density-profile for a fixed particle number}
In order to derive physical quantities for a fixed particle-number, we 
have to project the stationary
state onto a sector with a specified number of particles $M$. A projection
formalism for matrix-product states has been introduced recently in
Ref. \cite{Mallick}. We are now going to apply this formalism to the
present model with sequential dynamics (the case of sublattice-parallel
dynamics can be treated similarly). Because of particle-hole symmetry,
we restrict ourselves to less than half-filled systems where
$M \leq N = L/2$.

Let $\chi(m;\sigma_1,\ldots,\sigma_n)$ be a function which is $1$ if the
sites $1\ldots n$ are occupied by $m$ particles and
$0$ otherwise:
\begin{equation}
\label{ChiFunction}
\chi(m;\sigma_1,\ldots,\sigma_n) \;=\;
\delta \Bigl( n-m+\sum_{i=1}^n (\delta_{\sigma_i,3}-
\delta_{\sigma_i,0})\Bigr)\,.
\end{equation}
Then the probability to find the model with sequential updates and $M$
particles in the configuration $\{\sigma_1,\ldots,\sigma_N\}$ is given by
\begin{equation}
\label{SectorProbability}
P_M^{seq}(\sigma_1,\ldots.\sigma_N) \;=\;
\frac{1}{Z_{N,M}^{seq}} \,\, \chi(M; \sigma_1,\ldots,\sigma_N) \,\,
\langle W| \, A_{\sigma_1} \ldots A_{\sigma_N} \, |V \rangle\,,
\end{equation}
where $Z_{N,M}^{seq}$ is the normalization constant restricted to the
$M$-particle sector:
\begin{eqnarray}
\label{ConstrainedNormalization}
Z_{N,M}^{seq} &=& \langle W| \, G_{N,M} \, |V \rangle \\
G_{N,M} &=& \sum_{\sigma_1,\ldots,\sigma_N}
\chi(M;\sigma_1,\ldots,\sigma_N)\,
A_{\sigma_1} \ldots A_{\sigma_N}
\end{eqnarray}
The expression $G_{N,M}$ is the sum of all products of $N$ matrices
with $M$ particles. By definition, two of these objects can be combined
by convolution:
\begin{equation}
\label{Concatenate}
G_{N_1+N_2,M} \;=\; \sum_{j=0}^M \, G_{N_1,j}\,G_{N_2,M-j}\,.
\end{equation}
Physical quantities can be expressed in terms of combinations of these
objects. For example the density $c(x)$ to find a particle at site
$x$ in the original model on the ring with $L=2N$ sites and $M \leq N$
particles is given by
\begin{equation}
\label{ParticleDensity}
c(x) \;=\; \frac{1}{Z^{seq}_{N,M}}
\left\{
\begin{array}{ll}
\sum_{j=0}^{\min(x-1,M-1)} \,
\langle W| G_{x-1,j} \, A_1 \, G_{N-x,M-j-1} \, |V \rangle
& \mbox{if} \,\, x \leq L/2 \\[2mm]
\sum_{j=0}^{\min(L-x,M-1)} \,
\langle W| G_{L-x,j} \, A_2 \, G_{x-N-1,M-j-1} \, |V \rangle
& \mbox{if} \,\, x > L/2
\end{array}
\right.
\end{equation}
Similar expressions exist for the current and higher correlation functions.
Once $G_{n,m}$ is known, all these quantities can be computed
immediately.

\noindent
The expression $G_{n,m}$  can be defined recursively by
$G_{n,m}=0$ if $m<0$ or $m>n$,  $G_{0,0}=1$, and
\begin{equation}
G_{n,m} \;=\; A_0 \, G_{n-1,m} \,\,+\,\, (A_1+A_2)\,G_{n-1,m-1}
\,\,+\,\, A_3\,G_{n-1,m-2}\,.
\end{equation}
However, instead of solving this recurrence relation algebraically, it is much
simpler to use directly the representation (\ref{Representation}). We obtain
\begin{equation}
G_{n,m} =
\left(\begin{array}{cc}z_1&z_3-z_2-z_1\\0&z_2\end{array}\right)\,,
\end{equation}
where
\begin{equation}
z_1=\delta_{n,m}\,q^{-n}\,, \hspace{10mm}
z_2= \left( \hspace{-2mm}
\begin{array}{c} n \\ m \end{array} \hspace{-1.5mm}\right)
\,, \hspace{10mm}
z_3=2\,\sum_{j=0}^m\,\left( \hspace{-2mm}
\begin{array}{c} n \\ m-j \end{array} \hspace{-1.5mm}\right)\,
(1-q)^j q^{-j}\,.
\end{equation}
Using this result we obtain the normalization
(\ref{ConstrainedNormalization})
\begin{equation}
Z^{seq}_{N,M} \;=\;
2\,q^{-M} \, \sum_{j=0}^{M}\, \left( \hspace{-2mm}
\begin{array}{c} N \\ M-j \end{array} \hspace{-1.5mm}\right)\,
q^j \, q^{M-j}
\;=\;
2  q^{-M} \, (1-q)^{M-N} \, I_{1-q}(N-M,M+1)\,,
\end{equation}
where $I_z(n,m)$ is the regularized incomplete beta function. Now the
particle density profile (\ref{ParticleDensity}) can be computed easily.
Using that $A_1=\mbox{\bf 1}$ and $\langle W|G^{seq}_{n,m}A_2 =
\langle W|\delta_{n,m}(1-q)q^{-n-1}$ we get
\begin{equation}
\label{ExactDensity}
c(x) \;=\; \frac{1}{Z^{seq}_{N,M}}
\left\{
\begin{array}{ll}
Z^{seq}_{N-1,M-1}
& \mbox{if} \,\, x \leq L/2 \\[2mm]
(1-q)\,q^{x-2N-1}\, Z^{seq}_{x-N-1, M+x-2N-1}
& \mbox{if} \,\, x > L/2
\end{array}
\right.
\end{equation}
This formula holds for less than half filling $M \leq N=L/2$ (the case of
more than $L/2$ particles is related by particle-hole symmetry). Writing
$x=L-y+1$ and keeping the particle density $\rho=M/N \ll q$ fixed one can now 
derive the asymptotic expression for the particle
density in front of the defect in the thermodynamic limit:
\begin{equation}
\lim_{N \rightarrow \infty} \, 
c(L-y+1)  \;=\; (1-q) \, \Bigl( \frac{\rho}{q} \Bigr)^y
\end{equation}
A similar result was derived in Ref. \cite{Blockage} for sublattice-parallel
updating.
%
%
%
\subsection{Grand-canonical ensemble: Correlations for a large lattice}
Let us consider a grand-canonical ensemble of systems with $L \gg 1$
and an average particle number $\rho L$ where
$\rho$ is   some  given density.
Its probability distribution is given by 
Eq.~(\ref{SequMatrixProductComplete}) with the matrices~(\ref{Representation}).
The `fugacity'-parameter $\lambda$ has to be chosen such that the
particle-density is equal to $\rho\;$,
i.e. it has to solve the following equation:
\begin{eqnarray}
\label{rho-la}
2 N \rho \;=\;\frac{1}{Z_N^{seq}}\;\sum_{i=1}^{N}\;
\langle W|\,C^{i-1}\,(A_1+A_2+2A_3)\,C^{N-i}\,|V\rangle\,.
\end{eqnarray}
Like in the previous section we restrict to particle numbers less than $N=L/2$
(Results for systems with higher particle numbers are obtained exploiting
the particle-hole symmetry). Because of 
$C=A_0+A_1+A_2+A_3$ and $A_3=0$ Eq. (\ref{rho-la}) can be written as
\begin{eqnarray}
\label{rho-la-1}
1-2  \rho &=&\frac{1}{N\;Z_N^{seq}}\;\sum_{i=1}^{N}\;F_{N,i}\\
\label{rho-la-2}
\mbox{with}\;\;\;\;
F_{N,i}&=&\langle W|C^{i-1}A_0 C^{N-i}|V\rangle \nonumber
\;\;.
\end{eqnarray}
The computation of the matrix-elements $Z_N^{seq}$ and $F_{N,i}$
is done most easily in a representation where $C$ is diagonal.
Such a representation can be obtained by means of a similarity
transformation from the representation
(\ref{Representation}). We will use the following matrices and vectors
\begin{eqnarray}
\label{Representation-dia}
&&C=\left(\begin{array}{cc}q^{-1}&0\\0&1+\lambda\end{array}\right)\,, 
\hspace{7mm}
A_0=\lambda\left(\begin{array}{cc}0&\frac{2(q-1)}{q+\lambda q-1}\\
0&1 \end{array}\right)\,, 
\\
\label{Representation-dia1}
&&\langle W|=\Bigl(1,\frac{2 \lambda q}{q+\lambda q-1}\Bigr) \,, \hspace{15mm}
| V \rangle = \left( \hspace{-2mm}
\begin{array}{c} \frac{2(q-1)}{q+\lambda q-1}  \\[-0mm] 1 \end{array}
\hspace{-1.5mm}\right) 
\nonumber
\end{eqnarray}
to compute $Z_N^{seq}$ and $F_{N,i}$ as
\begin{eqnarray}
\label{Z-result}
Z_N^{seq}&=&\frac{2 }{q+\lambda q-1}\;\left\{\;
\lambda q \;(1+\lambda)^N\;+\;(q-1)\;q^{-N}\; \right\} \\
\label{g-result}
F_{N,i}&=&\frac{2 \lambda }{q+\lambda q-1}\;(1+\lambda)^{N-1}\;\left\{\;
\lambda\;q\;+\;
\frac{q-1}{(q(1+\lambda))^{i-1}}
\;\right\}
\;\;.\end{eqnarray}
The large-$N$ asymptotics of these expressions depend on the magnitudes
of the terms $1+\lambda$ and~$q^{-1}\;$.
Let us consider the case $1+\lambda>q^{-1}$ with $1+\lambda-q^{-1}=O(N^0)$
first. Inserting Eqs.~(\ref{Z-result}) and~(\ref{g-result})
into Eq.~(\ref{rho-la-1}) and approximating for $N \gg 1$
results in
\begin{equation}
\label{lambda-result}
\lambda=\frac{1-2 \rho}{2 \rho}
\;.
\end{equation}
This equation relates the parameter $\lambda$ to the density $\rho$
under the condition $1+\lambda>q^{-1}\;$, i.e. for
\begin{equation}
\label{condition}
\rho\;<\;\frac{q}{2}
\;.
\end{equation}
Eq. (\ref{lambda-result}) completes the description of the system for
the case $\rho<q/2\;$. One can now easily compute the density-profile
or correlation functions. Noticing that $C$ plays the role of a transfer matrix
one reads from Eq. (\ref{Representation-dia}) that all correlations decay
exponentially (see e.g. Ref. \cite{ReactionDiffusion})  with a correlation
length $\xi=\left\{ log (1+\lambda)q \right \}^{-1}\;$. Using the explicit
expression (\ref{lambda-result}) for $\lambda$ we find
\begin{equation}
\label{xi}
\xi \approx \left\{ log \frac{q}{2 \rho} \right \}^{-1}
\;.
\hspace{20mm}
(N \gg 1, \; \rho<q/2)
\end{equation}
This result is the same as the one obtained in Ref. \cite{Blockage} for
the model with sublattice-parallel dynamics.

In the case $\rho\ge q/2$ the above assumptions on
$\lambda$ do not lead to any result. In order to obtain a finite
density we have to assume $\lambda$ to be of the form
\begin{equation}
\label{lambda-form}
\lambda\;=\;-1\;+\;q^{-1}\;\left(\;1\;+\;\frac{\lambda'}{N} \right )
\end{equation}
with some number $\lambda'$ which is independent on $N\;$.
Computing again the large-$N$ asymptotics
of Eqs. (\ref{Z-result}) and (\ref{g-result}) and inserting it into
Eq. (\ref{rho-la-1}) results in the transcendental equation
\begin{equation}
\label{lambda-result-1}
\frac{1-2 \rho}{1-q}\;=\;\frac{ 1\;+\;(\lambda'-1)e^{\lambda'}}
{\lambda'\;\left(e^{\lambda'}-1 \right) }
\;\end{equation}
which is soluble for densities $\rho \ge q/2$.
Eqs. (\ref{lambda-form}) and (\ref{lambda-result-1})
relate the parameter $\lambda$ to the density $\rho\;$.
The correlation length, which is given by
$\xi= \left\{ log (1+\lambda)q \right \}^{-1}$
($\;\xi= - \left\{ log (1+\lambda)q \right \}^{-1}\;$ )
for $\lambda' > 0$ ($\;\lambda' < 0\;$) takes now the form
\begin{equation}
\label{xi1}
\xi \approx \frac{N}{|\lambda'|} \;,
\hspace{20mm}
(N \gg 1, \; \rho \ge q/2)
\end{equation}
i.e., it is proportional to the system size. Consequently,
correlations on a distance $k \ll N$ decay like $\lambda'\; k/N\;$.
This type of behavior is completely different from the one observed in
the $\rho < q/2$-phase. It indicates that the system behaves like  the one
undergoing sublattice-parallel dynamics  \cite{Blockage} where there is a
coexistence phase with non-exponential decay of correlations.
%
%
%
\section{Conclusion}
\label{ConclusionSection}
We have shown that MPS techniques can be applied successfully to an asymmetric
exclusion processes on a ring with a defect. In order to apply this method,
the model on a ring has to be reformulated as a four-state model on a linear
chain. This ansatz takes into account that translational invariance is 
broken by the defect while the system is still symmetric under
reflections with respect to the defect combined with particle-hole exchange.
The algebra we derived has very simple two-dimensional
representations which allow physical quantities like the density profile
to be computed directly. Another advantage of the matrix formalism is that
the model with different dynamical rules can be solved within the same
framework (i.e. with the same algebra and representations).

The ASEP discussed in this paper is a very simple one where particles
hop deterministically in the bulk. Our hope is that the present work
may show a way how to solve the more general problem with probabilistic
hopping in the bulk. This generalized model is controlled by three quantities,
the bulk hopping rate $p$, the defect hopping rate $q$, and the particle
density $\rho=M/L$. In the so-called Hamiltonian limit
$p,q \rightarrow 0$, $p/q=const$, it includes the case of
{\it random sequential updates} which is an outstanding problem.

When introducing a bulk hopping rate $p$, the algebra
(\ref{BulkEquations})-(\ref{RightBC}) has to be replaced by
the bulk equations
\begin{eqnarray}
[A_i,B_i]&=&0\hspace{17mm}(i=1\ldots 4) \nonumber \\
(1-p)A_iB_j&=&B_iA_j \hspace{10mm} (i,j)
\in \{(0,2),(1,0),(1,3),(3,2)\}\nonumber\\
A_jB_i+p\,A_iB_j&=&B_jA_i
\hspace{10mm} (i,j) \in \{(0,2),(1,0),(1,3),(3,2)\}\nonumber\\
(1-p)A_iB_j+p(1-p)A_1B_2&=&B_iA_j \hspace{10mm} (i,j) \in \{(0,3),(3,0)\}\\
(1-p)^2A_1B_2&=&B_1A_2\nonumber\\
A_2B_1+p A_0B_3 + p^2 A_1B_2+p A_3 B_0 &=& B_2A_1\nonumber
\end{eqnarray}
and the boundary conditions
\begin{eqnarray}
\langle W| B_0 &=& \langle W| A_0 \nonumber
\hspace{34mm}
A_0 |V\rangle \;=\; B_0|V\rangle \\
\langle W| (B_1+qB_2) &=& \langle W| A_1
\hspace{22mm}
(1-p)A_1|V\rangle \;=\; B_1 |V\rangle \\
\langle W| (1-q) B_2 &=& \langle W|A_2 \nonumber
\hspace{20mm}
(pA_1+A_2) |V\rangle \;=\; B_2 |V\rangle \\
\langle W| B_3 &=& \langle W| A_3 \nonumber
\hspace{34mm}
A_3|V\rangle  \;=\; B_3 |V\rangle
\end{eqnarray}
This algebra is even more complicated as the previous one and we were not
able to find representations or to prove its consistency. 
However, following the ideas of \cite{Rajewsky}, 
one could again assume the additional relations given in
Eq. (\ref{SimplificationAnsatz}) hold. This reduces the algebra to
\begin{eqnarray}
&& pA_1A_0 = A_0 \,,\hspace{12mm}
pA_1A_3 = A_3 \,,\hspace{12mm}
pA_0A_2 = (1-p)A_0 \,,\hspace{12mm}
pA_3A_2 = (1-p)A_3 \nonumber\,, \\
\label{Problem}
&&
A_0A_3=A_3A_0 =\frac{1-p}{p(2-p)}(A_1+A_2) \,,\hspace{8mm}
A_1A_2 = \frac{(1-p)^2}{p(2-p)}A_1 + \frac{1}{p(2-p)}A_2 \,,
\end{eqnarray}
together with the boundary equations
\begin{equation}
\langle W|A_2=\frac{1-q}{q}\langle W|\,,
\hspace{20mm}
A_1|V\rangle = \frac{1}{p}|V\rangle\,.
\end{equation}
Likewise, although this algebra is much simpler, we were not able to prove
its consistency and the existence nontrivial representations. 
As usual in ASEPs including a Hamiltonian limit,
such nontrivial representations are expected to be infinite dimensional.
The ASEP with a defect and random sequential updates remains as an
open problem.
%
%
%
%
\\[10mm]
\noindent
{\bf Acknowledgments}\\[1mm]
We would like to thank  I. Peschel and in particular A. Honecker 
for useful hints and interesting discussions. 
S.S. greatfully acknowledges financial support by 
the Deutsche Forschungsgemeinschaft. H.H. would like to thank the Minerva 
foundation for financial support.
%
%
%
%
%

\end{document}